\documentstyle[12pt,epsfig]{article}
\newcommand{\cl}{\centerline}
\def\beq{\begin{equation}}
\def\eeq{\end{equation}}
\def\bea{\begin{eqnarray}}
\def\eea{\end{eqnarray}}
\def\bq{\begin{quote}}
\def\eq{\end{quote}}

\def\PLB{{\it Phys. Lett.} }
\def\PRL{{\it Phys. Rev. Lett.} }
\def\NP{{\it Nucl. Phys.} }
\def\PR{{\it Phys. Rev.} }

\def\JP{{\it J. Phys.} }

\def\ZPC{{\it Z. Phys. } }
\def\+{\;\;\;\;+\;\;\;\;}
\def\-{\;\;\;\;-\;\;\;\;}
\def\({\left(\;\;\;\;}
\def\){\;\;\;\;\;\;\right)}
\def\={\;\;\;\;=\;\;\;\;}

\def\bef{\begin{figure}}
\def\befa{\begin{figure}}
\def\enf#1{\label{#1}\end{figure}}
\def\bc{\begin{center}}
\def\ec{\end{center}}
\def\bei{\begin{itemize}}
\def\ei{\end{itemize}} 
\def\be{\begin{eqnarray*}}
\def\ed{\end{eqnarray*}}
\def\bee{\begin{eqnarray}}
\def\eee{\end{eqnarray}}
\def\ben{\begin{equation}}
\def\edn#1{\label{#1}\end{equation}}
\def\bef#1#2{\begin{fmfgraph*}(#1,#2)}
\def\enf{\end{fmfgraph*}}
\def\nn{\nonumber\\}

\def\fr#1#2{\frac{#1}{#2}}
\def\bn{\bar{n}}

\def\jpg1#1#2#3{J.Phys.{\bf G}: Nucl. Part. Phys. {\bf #1},#2(#3)}

\def\g5{\gamma_5}

\def\s#1{\slash\!\!\!{#1}}

\def\->#1{\overrightarrow{#1}}
\def\<-#1{\overleftarrow{#1}}
\def\v0{\->{0}}

\def\nb{\bar{\nu}}

\def\yref#1{eq.~(\ref{#1})} 
\def\gappeq{\mathrel{\rlap {\raise.5ex\hbox{$>$}}
{\lower.5ex\hbox{$\sim$}}}}

\def\lappeq{\mathrel{\rlap{\raise.5ex\hbox{$<$}}
{\lower.5ex\hbox{$\sim$}}}}

\begin{document}
\topmargin -2.0cm
\oddsidemargin -0.3cm
\evensidemargin -0.8cm
\pagestyle{empty}
\begin{center}
{\bf PQCD ANALYSIS OF INCLUSIVE SEMILEPTONIC DECAYS OF A POLARIZED 
$\Lambda_b$ BARYON}\\
\vskip 1.0cm
\cl{Tsung-Wen Yeh,}
\vskip 0.3cm
\cl{Institute of Physics,}\par
\cl{Academia Sinica,}\par
\cl{Taipei, Taiwan, R.O.C.}\par
\vspace*{0.75cm}
{\it \today}\\

\vspace*{1.0cm}
{\bf Abstract}
\vskip 0.3cm
\end{center}

We investigate the $\Lambda_b$ polarization problem in the
inclusive semileptonic decays of a polarized $\Lambda_b$ baryon, using the 
modified perturbative QCD formalism which includes  
Sudakov suppression. 
According to HQEFT, we show that, at the leading
order in the $1/M_b$ expansion, the polarized and unpolarized
distribution functions become one single universal distribution function. 
To explore the mechanisms which determine the spin properties of
a polarized $\Lambda_b$ baryon, we construct   
four formalisms which are the naive quark model (QM),
the modified quark model (MQM)
, the naive parton model (PM), and the modified
parton model (MPM), and calculate their
corresponding $\Lambda_b$ polarizations, denoted as $\rm P$'s.
The modified quark and parton models are with Sudakov suppression.
The resulting $\rm P$'s   
are -0.23 (QM), -0.94 (MQM), -0.37 (PM), and -0.68 (MPM), respectively. 
We note that $\rm P_{MQM}$ (equal to -0.94 )
is very close to the $b$ quark polarization asymmetry,
$A_{RL}=-0.94$,
calculated at the $Z$ vertex in $Z \to b \bar{b}$ process, 
and that $\rm P_{MPM}$ (equal to -0.68 ) 
is also very close to the $\Lambda_b$ baryon polarization 
(equal to -0.68 ) which was estimated from
the fragmentation processes under the heavy quark limit.  
Based on our analysis, there exists no any paradox in the
theoretical explanations of the $\Lambda_b$ polarization 
for the experimental data.
\newpage
\setcounter{page}{1}
\pagestyle{plain}

\section{Introduction}
\hskip 0.6cm
A recent measurement by ALEPH Collaboration \cite{ALEPH} indicated that 
the $\Lambda_b$ polarization was -0.23. This deviated largely 
from the Standard Model expectation that the $\Lambda_b$ polarization
should be -0.94 \cite{CLOSE,Mannel-Schular,KPTung}. 
It was, therefore,  argued that the spin-spin interactions 
between the $b$ quark and the light quarks and gluons produced
from the vacuum as the $b$ quark undergoing its fragmentation,
should bring about large spin-flip of the $b$
quark spin.
A heavy quark effective field theory (HQEFT) calculation 
showed that the final $\Lambda_b$ polarization was -0.68\cite{FP}.     
   
In this paper, we would like to re-interpret the measurement 
by the ALEPH Collaboration. The $\Lambda_b$ polarization, denoted as $\rm P$,
could be related to the experimental measured quantity, 
$R=1.12\pm0.10$,  as
\bee
R=\fr{<E_\nu^*>(<E_l^*>+<P_l^*({\rm P})>)}
{<E_l^*>(<E_\nu^*>+<P_\nu^*({\rm P})>)}\;\;. 
\label{P-R}
\eee 
Based on the naive quark model, the averaged quantities, $<E_l^*>$, $<E_\nu^*>$,
$<P_l^*>$ and $<P_\nu^*>$ were calculated in the $\Lambda_b$ rest frame 
. The resulting $\rm P$ was equal to $-0.23$ \cite{ALEPH}.
Since the spectra are still not available, the 
explanation for the $\Lambda_b$ polarization 
is therefore theoretical dependent. 
For the purpose to explore the mechanisms which control 
the spin properties of a polarized $\Lambda_b$ baryon,
we shall construct four formalisms to investigate, 
which are the naive quark model (QM), 
the modified quark model (MQM), the naive parton model (PM), 
and the modified parton model (MPM). 
The modified quark model and modified parton model 
are with Sudakov suppression.  

We emphasize the importance of the transverse degrees of freedom of partons
inside a $\Lambda_b$ baryon in our analysis. 
First of all, the transverse momenta regularize the divergences 
when the outgoing $q$ quark is approaching the end point \cite{LIYU1}. 
And second, the transverse momenta enhance the contributions from the 
longitudinal component of the $\Lambda_b$ baryon spin vector.   
These effects indicate that the intrinsic $b$ dependence of  
the distribution function, $e.g.$, $f(z,b)$, is nonignorable.    
The form of the $b$ dependence of the distribution function
could be determined by exploiting the  
infrared (IR) renormalon contributions  
\cite{LIYU2,K-STERMAN}, such that one can parameterize  
$f(z,b)$ as $\exp{[-tM^2b^2]} f(z)$. 
The parameter $t$ will be determined from the charged lepton spectra.   

The arrangement of our paper is as follows. In the next section,
we develop a power
expansion scheme which is appropriate for heavy quark system.
By employing the HQEFT,
we generalize the naive collinear expansion 
scheme \cite{QIU} to include heavy massive quark partons and
to apply to decay processes.
Using this generalized collinear expansion scheme, we then show that, 
in the heavy quark limit, 
there exists an universal distribution function
which respects both the unpolarized and polarized matrix elements.
By taking into account the radiative corrections,
the factorization formula is expressed as the convolution of a hard scattering
amplitude with a jet and a universal soft function. 
In Section 3, we construct the four formalisms based on the
factorization formula.
Section 4 is the numerical result and Section 5 the conclusion.
An Appendix is presented for those details skipped by the main text.

\section{Factorization Formula }

The quadruple differential decay rate for polarized 
${\Lambda_b}\to X_q\ell\nb$ is expressed as
\bee
\fr{d^4\Gamma}{dE_l dq^2 dq_0 d\cos\theta_l}
&=&\fr{|V_{qb}|^2{G_F^2}}{256\pi^4M}L^{\mu\nu}W_{\mu\nu}\;\;, 
\eee
where $M$ is the $\Lambda_b$ baryon mass,  
$L^{\mu\nu}$ is the leptonic tensor  
and $W_{\mu\nu}$ is the hadronic tensor.  
 
The kinematically independent variables $E_l$, $q$, $q_0$ and $\cos\theta_l$
are expressed as follows. We set our working frame in the $\Lambda_b$ baryon
rest frame and specify relevant momenta as 
\bee
P=\fr{M}{\sqrt{2}}(1,1,{\bf 0}), p_l=(p_l^+,0,{\bf 0}), 
p_{\nb}=(p_{\nb}^+,p_{\nb}^-,{\bf p}_{\nb\perp})\;.
\eee
$E_l$, $q$, and $q_0$ are expressed as $E_l=p_l^+/\sqrt{2}$, 
$q^2=2p_l^+ p_{\nb}^-$, and $q_0=(p_l^+ + p_{\nb}^+ + p_{\nb}^-)/\sqrt{2}$,
respectively. We let $P_b=P-l$ whose square is set as $P_b^2\approx M_b^2$ 
, $M_b$ the $b$ quark mass. $l$ is the momentum of the light degree of freedom
inside the $\Lambda_b$ baryon, 
and has a large plus component and small transverse
components ${\bf l}_{\perp}$. The final state quark momentum is 
$P_q=P-l-q$. $\theta_l$ is the angle between the third component of $p_l$
and that of the $b$ quark polarization vector, 
$S_b=(S_b^+,S_b^-,{\bf S}_{b\perp})$.

It is convenient to use dimensionless variables $x=2E_l/M$, $y=q^2/M^2$,
and $y_0=2q_0/M$.
The integration regions for $x$, $y$ and $y_0$ are 
\bee
0\le x\le 1, \hspace{1cm} 0\le y \le x,\hspace{1cm} \fr{y}{x}+x\le y_0\le 1+y.
\eee
 Note that we have chosen $M$ as scale variable and
have set $m_q=0$ for simplicity.
The cases for $m_q\ne 0$ will be considered in our future publish. 


By optical theorem, $W^{\mu\nu}$ relates to the forward 
matrix element $T^{\mu\nu}$ as 
\bee
W^{\mu\nu}=-\frac{\mbox{Im}(T^{\mu\nu})}{\pi}.
\eee
The lowest order of $T^{\mu\nu}$ is defined as
\bee
T^{\mu\nu}(P,q,S)&=&-i\int d^4 y e^{iq\cdot y}
\langle \Lambda_b(P,S)|{\cal T}[
J^{\dag\mu}(0),J^{\nu}(y)]|\Lambda_b(P,S)\rangle \nn  
&=&-i\int \fr{d^4P_b}{(2\pi)^4} S^{\mu\nu}(P_b-q)T(P,S,P_b),
\label{st}
\eee
where $S^{\mu\nu}(P_b-q)$ describes the short distance $b\to W q$ 
decay subprocess
and $T(P,S,P_b)$ relates the corresponding 
long distance matrix element  
\bee
T(P,S,P_b)=\int d^4 y e^{iP_b\cdot y}\langle \Lambda_b(P,S)|\bar{b}(0)b(y)
|\Lambda_b(P,S)\rangle.  
\label{t1}
\eee
$J^{\mu}=\bar{q}\gamma^{\mu}(1-\gamma_5)b$ is the V $-$ A current.
Because the momentum $P_b$ in $S^{\mu\nu}(P_b)$ has non-collinear
component which would give high order power contributions \cite{QIU},
it is then necessary to investigate the collinear expansion for
$T^{\mu\nu}$.  The main procedures are demonstrated as follows.  

For a $\Lambda_b$ baryon which carries momentum $P$ and mass $M$, 
we specify $P$ to be  
\bee
P^{\mu} &=& p^{\mu} + \fr{M^2}{2 p\cdot n}n^{\mu}, 
\eee
where $p^2=n^2=0$, $p\cdot n=P\cdot n$. 
The $b$ quark, inside the $\Lambda_b$ baryon,
 carries momentum $P_b$ chosen as
\bee
P_b^{\mu}&=& z {p}^{\mu} 
+ \fr{P_b^2 + P_{g\perp}^2}{2 P_b\cdot n }
 n^{\mu} + P_{b\perp}^{\mu} \\ 
&=& \hat{P}_b^{\mu} + \fr{P_b^2 - M_{b}^2}{2 P_b\cdot n }
n^{\mu}, 
\eee  
where $\hat{P}_b^2=M_b^2$ is the on-shell part of $P_b$
and the momentum fraction $z$ defined by $z=P^+_b/P^+=1-l^+/P^+$.
By the parameterization of $P_b$, the 
$b$ quark propagator is then expressed as 
\bee
\fr{i}{\s{P}_b -M_b + i\epsilon}=\fr{i(\hat{\s{P}}_b + M_b)}
{\s{P}_b -M_b + i\epsilon}+ \fr{i\s{n}}{2 P_b\cdot n }\;\;.
\label{heavyprop}
\eee
The second part of \yref{heavyprop} is called special propagator
introduced by Qiu \cite{QIU}. The special propagator
describes the short distance nature of the loop propagator in a
Feynman diagram.   
To generalize the naive collinear expansion scheme to include heavy
massive parton,
we employ the technology of the HQEFT to rescale the $b$ quark field, 
$b(x)$, as $b_v(x)=\exp{(iM_b v\cdot x)}\fr{1+\s{v}}{2}b(x)$. In this
way, $P_b$ is parameterized as $P_b=M_b v +k$, with $k$ the residual
momentum. The rescaled $b$ quark field, $b_v(x)$, carries the
residual momentum $k$ and has a small effective mass $\bar{\Lambda}$,
with $\bar{\Lambda}\equiv M-M_b$. 
Since $k$ is of order $O(\Lambda_{QCD})$ and $\bar{\Lambda}$ 
much smaller than $M_b$,
it is thus expected that 
the whole program of Qiu's collinear expansion for massless parton
would be applicable for $b_v(x)$.  
We now demonstrate the main procedure of this generalized collinear
expansion.
First, one expands $k$ as $k=\xi p + (k-\xi p)$,
where $\xi p = (z - 1) p + \bar{\Lambda}/M p$.
Secondly, one performs a Taylor expansion of $S^{\mu\nu}(k)$  around 
$S^{\mu\nu}(\xi p)$ as 
\bee
S^{\mu\nu}(k)=S^{\mu\nu}(\xi p) + \fr{\partial S^{\mu\nu}}{\partial k^{\alpha}}
|_{k=\xi p} (k-\xi p)^{\alpha} + \cdots \;. 
\label{smunu}
\eee
The high order terms \yref{smunu} are
to replace the higher order gluon field operators in the higher
order components of $T^{\mu\nu}$ by the relating covariant operators. 
In the same way, the non-collinear component of $k$ in $T(k)$ would
yield high order contributions. It happens when $T(k)$ is contracted
with $\s{p}$ ( or $\s{p}\gamma_5$).  
As a result, at leading order, we could write $T^{\mu\nu}$ in the form 
\bee
T^{\mu\nu}(P,q,S)
&\approx& -i\int \fr{d^4 k}{(2\pi)^4} \left\{ 
[S^{\mu\nu}(k=\xi p,q)\s{P}_b][T(P,S,k=\xi p)\fr{\s{n}}{4P_b\cdot n}]\right. \nn
&&\hspace{1cm}\left. -[S^{\mu\nu}(k=\xi p,q)\s{S}_b\gamma_5]
[T(P,S,k=\xi p)\fr{\s{n}\gamma_5}{4S_b\cdot n}]\right\}\;\;.
\label{exp1}
\eee
(The details of this proof will be described in the Appendix A.)

We now follow \cite{LIYU1} to derive the factorization formula 
for the inclusive semileptonic decay $\Lambda_b\to X_q\ell\nu$. 
After including the radiative corrections, the formula is written as
\begin{eqnarray}
\frac{1}{\Gamma^{(0)}}\frac{d^3\Gamma}{dxdydy_0d\cos\theta_l}
&=&\fr{M^2}{2}\int^{z_{\rm max}}_{z_{\rm min}}{dz} \int d^2{\bf l_\bot}
\nonumber \\
&&\times
S(z,{\bf l_\bot},\mu)J(z,P_q^-,{\bf l_\bot},\mu)
H(z,P_{q}^-,\mu)\;,
\label{as}
\end{eqnarray}
with $\Gamma^{(0)}=\frac{G_F^2}{16\pi^3}|V_{qb}|^2{M}^5$ 
and $\mu$ the renormalization and factorization scale. 
The transverse momentum ${\bf l}_\bot$ has been 
introduced for the regularization
of the end point singularities \cite{LIYU1}. 
In order to make factorization of the intrinsic 
transverse momentum from the radiative transverse momentum of
$J$, we transform \yref{as} into the impact $\bf b$ space   
\bee 
\frac{1}{\Gamma^{(0)}}\frac{d^3\Gamma}{dxdydy_0d\cos\theta_l}
&=&\fr{M^2}{2}\int^{1}_x{dz} \int \frac{d^2{\bf b}}{(2\pi)^2}
\nonumber \\
&&\times
{\tilde S}(z,{\bf b},\mu){\tilde J}(z,P_q^-,{\bf b},\mu)
H(z,P_{q}^-,\mu)\;.  
\label{asb}
\eee

To deal with the collinear and soft divergences resulting from
the radiative corrections for massless parton inside the jet, 
the resummation technique is necessary  
and these divergences could be resummed into a Sudakov form factor
\cite{LIYU1}.
The jet function is then re-expressed in the form 
\bee
{\tilde J}(z,P_q^-,b,\mu)=\exp{[-2s(P_q^-,b)]}{\tilde J}(z,b,\mu),
\label{js}
\eee 
where $\exp{[-2s(P_q^-,b)]}$ is the Sudakov form factor. 
 
The scale invariance of the differential decay rate in \yref{asb}
and the Sudakov form factor in \yref{js}
requires the functions $\tilde J$, $\tilde S$, and $H$ to obey the following
RG equations:
\begin{eqnarray}
{\cal D}{\tilde J}(b,\mu)&=& -2\gamma_q {\tilde J}(b,\mu)\;,
\nonumber \\
{\cal D}{\tilde S}(b,\mu)&=& -\gamma_S{\tilde S}(b,\mu)\;,
\nonumber \\
{\cal D}H(P_q^-,\mu)&=& (2\gamma_q+\gamma_S)H(P_q^-,\mu)\;,
\label{ter}
\end{eqnarray}
with
\begin{equation}
{\cal D}=\mu\frac{\partial}{\partial\mu}+\beta(g)
\frac{\partial}{\partial g}\;.
\end{equation}
$\gamma_q=-\alpha_s/\pi$ is the quark anomalous dimension in axial
gauge, and $\gamma_S=-(\alpha_s/\pi)C_F$ is the anomalous dimension of
$\tilde S$.
After solving \yref{ter}, we obtain the evolution of all the convolution
factors in \yref{asb},
\begin{eqnarray}
{\tilde J}(z,P_q^-,b,\mu)&=& {\rm exp}\left[-2s(P_q^-,b)-2\int^\mu_{1/b}
\frac{d{\bar\mu}}{{\bar\mu}}\gamma_q(\alpha_s(\bar\mu))\right]
{\tilde J}(z,b,1/b)\;,
\nonumber\\
{\tilde S}(z,b,\mu)&=& {\rm exp}\left[-\int^\mu_{1/b}
\frac{d{\bar\mu}}{{\bar\mu}}\gamma_S(\alpha_s(\bar\mu))\right]f(z,b,1/b)\;,
\nonumber\\
H(z,P_q^-,\mu)&=& {\rm exp}\left[-\int^{P_q^-}_\mu
\frac{d{\bar\mu}}{{\bar\mu}}[2\gamma_q(\alpha_s(\bar\mu))
+\gamma_S(\alpha_s(\bar\mu))]\right]H(z,P_q^-,P_q^-)\;.
\label{un}
\end{eqnarray}
In the above solutions, we set the $1/b$ as an IR cut-off for
single logarithism evolution. However, the intrinsic $b$ dependence of $f(z,b)$
gives more nonperturbative higher order contributions which
could be determined by exploiting the IR renormalon contributions. 
We employ a minima setting for $f(z,b)$ as
\bee
f(z,b)=f(z)e^{-\Sigma(b)}\;\;.  
\eee
The IR renormalon contributions arising from a real soft gluon 
attaching the two valence $b$ quarks as one calculating the
Sudakov form factors in \yref{js}. As a result, 
one could simply parameterize $\exp{[-\Sigma(b)]}$ in the form  
\bee
e^{-\Sigma(b)}=e^{-tM^2b^2}
\eee  
\cite{LIYU2}.
A few comments are needed. For the end point regime where the 
Sudakov suppression dominates, we employ the
approximation 
\bee
f(z,b)=f(z)
\label{suda}
\eee
, while for other regimes which are not under the control of the
Sudakov suppression, we take into account 
the intrinsic $b$ dependence of $f(z,b)$ to give more suppressions  
\bee
f(z,b)=e^{-tM^2b^2}f(z)\;\;.  
\label{renor}
\eee  
We make further approximations such that $f(z,b,1/b)= f(z,b)$,
${\tilde J}(z,b,1/b)={\tilde J}^{(0)}(z,b)$, and
$H(z,P_q^-,P_q^-)=H^{(0)}(z,P_q^-)$. 

Combining the above results, 
we write factorization formula for the inclusive
 semileptonic $\Lambda_b$ baryon decay as
\begin{eqnarray}
\frac{1}{\Gamma^{(0)}}\frac{d^4\Gamma}{dxdydy_0d\cos\theta_l}
&=&M^2\int^{1}_{x}
dz\int_0^{\infty}\frac{bdb}{4\pi}{\tilde J}^{(0)}(z,b)
H^{(0)}(z,P_{q}^-) e^{-S(P_q^-,b)} 
\nonumber \\
&&\times
\left\{
\begin{array}{cc}
f(z)& \makebox{for $x$ in end point regimes}\;, \\
\exp[-tM^2b^2]f(z)& \makebox{for $x$ in other regimes}\;,
\end{array}
\right.
\label{as1}
\end{eqnarray}
where 
\bee
S(P^-_q,b)=2 s(P^-_q,b)-\int^{P^-_q}_{1/b}\fr{d\bar{\mu}}{\bar{\mu}}
[2\gamma_q(\bar{\mu})+\gamma_S(\bar{\mu})]\;.
\eee
The parameter $t$ will be determined by interpolating from the
end point regimes for the relevant charged lepton spectrum.

Let's now discuss how to parameterize $T(k)$ defined in \yref{t1}.
As discussed in previous paragraphs that,
at leading order, $T(k)$ should not be contracted with $\s{p}
$ (or $\s{p}\gamma_5)$. 
So, we recast $T(k)$ in the form 
\bee
T(k)=\fr{1}{4}P_b\cdot n f(z) \s{\bn} 
- \fr{1}{4} S_b\cdot n g(z) \s{\bn}\gamma_5 
\eee
where $\bn^{\mu}=p^{\mu}/|p^{\mu}|$.
The unpolarized and polarized distribution functions, $f(z)$ and $g(z)$, are 
defined as
\bee
f(z)&=&\fr{1}{P_b\cdot n}\int \fr{d\lambda}{2\pi}
e^{-i\xi\lambda n} 
\langle P,S|\bar{b}_v(0)\s{n}b_v(\fr{\lambda n}{P\cdot n})|P,S\rangle
\eee
and
\bee
g(z)&=&\fr{1}{S_b\cdot n}\int\fr{d\lambda}{2\pi}
e^{-i\xi\lambda n} 
\langle P,S|\bar{b}_v(0)\s{n}\gamma_5b_v(\fr{\lambda n}{P\cdot n})
|P,S\rangle\;. 
\eee
It is easy to show that $f(z)$ and $g(z)$, in the heavy quark limit,
share a common matrix element which could be described by an universal
distribution function, $f_{\Lambda_b}(z)$. This just reflects the heavy
quark spin symmetry.   
We adopt the distribution function proposed in \cite{LIYU1} in the form 
\bee
f_{\Lambda_b}(z)=\fr{N z^2(1-z)^2}{((z-a)^2 + \epsilon z)^2}\theta(1-z)\;. 
\eee 
The parameters $N$, $a$ and $\epsilon$ are fixed by first 
three moments 
\begin{eqnarray}
& &\int_0^1 f_{\Lambda_b}(z) dz=1\;,
\nn
& &\int^1_0 dz (1-z) f_{\Lambda_b}(z)={\bar \Lambda}/M +{\cal O}
(\Lambda^2_{\rm QCD}/M^2)\;,
\nn 
& &\int^1_0 dz (1-z)^2 f_{\Lambda_b}(z)=\frac{\bar \Lambda^2}{M^2} +\frac{2}{3}
K_b+{\cal O} (\Lambda^3_{\rm QCD}/M^3)\;.
\label{eq31}
\end{eqnarray}
By substituting these constants, 
\bee
M=5.641 {\rm GeV}\;,\;\;\;\;M_b=4.776 {\rm GeV}\;,\;\;\;\; K_b=0.012\pm
0.0026\;,
\label{con}
\eee
into \yref{eq31}, we determine the parameters $N$, $a$ and $\epsilon$ to be 
\begin{equation}
N=0.10615\;,\;\;\;\; a=1\;,\;\;\;\; \epsilon=0.00413\;.
\label{eq55}
\end{equation}
For simplicity we will omit the subscript of $f_{\Lambda_b}(z)$ 
in the following text.

\section{Differential Decay Rates} 

In this section, we employ the factorization formula 
\yref{as1} to construct four formalisms which are 
the naive quark model (QM), 
the modified quark model (MQM),
the naive parton model (PM), 
and the modified parton model (MPM).
The charged lepton spectrum for the decay
$\Lambda_b\to X_q \ell\nu$ from the naive quark model
are expressed as
\bee
\fr{1}{\Gamma^{(0)}}\fr{d^2 \Gamma^{\rm T}_{\rm QM}}{dxd\cos\theta_\ell}
&=&\fr{1}{\Gamma^{(0)}}\fr{d^2 \Gamma^{\rm U}_{\rm QM}}{dxd\cos\theta_\ell}
+{\rm P}_{\rm QM}\cos\theta_\ell
\fr{1}{\Gamma^{(0)}}\fr{d^2 \Gamma^{\rm S}_{\rm QM}}
{dxd\cos\theta_\ell} \;, 
\label{eq53}
\eee
with
\bee
\fr{1}{\Gamma^{(0)}}\fr{d^2 \Gamma^{\rm U}_{\rm QM}}{dxd\cos\theta_\ell}
= \fr{x^2}{6}(3-2x)\;, 
\eee
and
\bee
\fr{1}{\Gamma^{(0)}}\fr{d^2 \Gamma^{\rm S}_{\rm QM}}{dxd\cos\theta_\ell}
=\fr{x^2}{6}(1-2x)\;.  
\eee
We draw the unpolarized, polarized and total charged leptonic spectra 
of QM in Fig.1 and denote them as curve 1. 

By taking into account Sudakov suppression from the resummation of
large radiative corrections, and substituting $f(z,b)=\delta (1-z)
\exp{[-t M^2b^2]}$,
$H^{(0)}=(x-y)[(y_0-x)+\rm P_{MQM}\cos\theta_\ell(y_0-x-2y/x)]$ and 
the Fourier transform of $J^{(0)}=\delta(P_q^2)$
with $P_q^2=M^2(1-y_0+y-p_{\bot}^2/M_B^2)$ into eq.~(\ref{as1}),
we derive the modified quark model spectrum. This spectrum is,
after integrating eq.~(\ref{as1}) over $z$ and $y_0$, described by
\bee
\frac{1}{{\Gamma}^{(0)}}\frac{d^2 \Gamma^{\rm T}_{\rm MQM}}{dxd\cos\theta_\ell}=
\frac{1}{{\Gamma}^{(0)}}\frac{d^2 \Gamma_{\rm MQM}^{\rm U}}{dxd\cos\theta_\ell}+
{\rm P_{MQM}}\cos\theta_\ell
\frac{1}{{\Gamma}^{(0)}}\frac{d^2 \Gamma_{\rm MQM}^{\rm S}}{dxd\cos\theta_\ell}
\;\;,
\label{MQM0}
\eee
\begin{eqnarray}
\frac{1}{{\Gamma}^{(0)}}\frac{d^2 \Gamma_{\rm MQM}^{\rm U}}{dxd\cos\theta_\ell}
&=&M \int_0^{x}dy \int_0^{1/\Lambda}db e^{[-t^{\rm U} M^2 b^2]}
 e^{-S(P_q^-,b)}  (x-y) \eta
\nonumber \\
&& \times\left[(1+y-x) J_1(\eta M b)
-\frac{2}{M b} \eta J_2(\eta M b) +\eta^2 J_3 (\eta M b)\right]\;,
\label{MQMu}
\end{eqnarray}
\begin{eqnarray}
\hspace{-0.5cm}\frac{1}{{\Gamma}^{(0)}}
\frac{d^2 \Gamma_{\rm MQM}^{\rm S}}{dxd\cos\theta_\ell}
&=&M \int_0^{x}dy \int_0^{1/\Lambda}db e^{[-t^{\rm S} M^2 b^2]} e^{-S(P_q^-,b)}  (x-y) \eta
\nonumber \\
&& \times\left[(1+y-x-2\fr{y}{x}) J_1(\eta M b)
-\frac{2}{M b} \eta J_2(\eta M b) +\eta^2 J_3 (\eta M b)\right]\;,
\label{MQMp}
\end{eqnarray}
where $P_q^-=(1-y/x)M/\sqrt{2}$, $\eta=\sqrt{(x-y)(1/x-1)}$ and
$J_1$,$J_2$,$J_3$ are the Bessel functions of order 1, 2 and 3, respectively.
The parameters $t^{\rm U}$ and $t^{\rm S}$ are chosen as    
$t^{\rm U}=0.005$ and $t^{\rm S}=0.0423$.
Note that we have made an approximation by substituting 
$f(z)\exp{[-t^{\rm U, S} M^2 b^2]}$ for the end point regimes. 
The spectra of MQM are drawn in Fig.1 and denoted as curve 2.
One observes that the polarized and total spectra of MQM deviate
largely from those of QM. The profile of the total spectrum  of MQM
is broader than that of QM and the peak position sifts to near 
$x=1$ end point. These differences between the MQM spectra and
the QM spectra indicate the resulting $\Lambda_b$ polarizations 
to be largely different, which are equal to  -0.94 and -0.23
for ${\rm P_{MQM}}$ and ${\rm P_{QM}}$ respectively.  

The naive parton model spectra are obtained
by adopting $H^{(0)}=(x-y)[(y_0-x-(1-z)y/x)+
{\rm P_{PM}}\cos\theta_\ell(y_0-x-(1+z)y/x)]$ and
$P_q^2=M^2[1-y_0+y-(1-z)(1-y/x)]$. With integration over $y_0$, we derive
\beq
\frac{1}{{\Gamma}^{(0)}}\frac{d^2 \Gamma_{\rm PM}}{d xd\cos\theta_\ell}=
\int_0^{x}dy \int_x^{1} dz f(z) (x-y)[(y+z-x)+{\rm P_{PM}}\cos\theta_\ell
(y+z-x-2z\fr{y}{x})]\;.
\label{e55a}
\eeq
The spectra are shown as curve 3 in Fig.1. 

We finally come to the charged lepton spectra of the modified parton model 
that takes into account both large perturbative and nonperturbative
corrections,
\bee
\frac{1}{{\Gamma}^{(0)}}\frac{d^2 \Gamma^{\rm T}_{\rm MPM}}{dxd\cos\theta_\ell}=
\frac{1}{{\Gamma}^{(0)}}\frac{d^2 \Gamma_{\rm MPM}^{\rm U}}{dxd\cos\theta_\ell}+
{\rm P_{MPM}}\cos\theta_\ell
\frac{1}{{\Gamma}^{(0)}}\frac{d^2 \Gamma_{\rm MPM}^{\rm S}}{dxd\cos\theta_\ell}
\;\;,
\label{MPM0}
\eee
\begin{eqnarray}
\frac{1}{{\Gamma}^{(0)}}\frac{d^2 \Gamma_{\rm MPM}^{\rm U}}{dxd\cos\theta_\ell}
&=&M \int_0^{x}dy \int_0^{1/\Lambda}db e^{-S(P_q^-,b)}
e^{-t^{\rm U} M^2 b^2} f(z)  (x-y) \eta
\nonumber \\
&& \times\left[(1+y-x) J_1(\eta M b)
-\frac{2}{M b} \eta J_2(\eta M b) +\eta^2 J_3 (\eta M b)\right]\;,
\label{MPMu}
\end{eqnarray}
\begin{eqnarray}
\hspace{-0.5cm}\frac{1}{{\Gamma}^{(0)}}
\frac{d^2 \Gamma_{\rm MPM}^{\rm S}}{dxd\cos\theta_\ell}
&=&M \int_0^{x}dy \int_0^{1/\Lambda}db e^{-S(P_q^-,b)}e^{-t^{\rm S} M^2 b^2}
f(z)  (x-y) \eta
\nonumber \\
&& \times\left[(z+y-x-2z\fr{y}{x}) J_1(\eta M b)
-\frac{2}{M b} \eta J_2(\eta M b) +\eta^2 J_3 (\eta M b)\right],
\label{MPMp}
\end{eqnarray}
with $\eta=\sqrt{(x-y)(z/x-1)}$. 
The parameters $t^{\rm U,S}$ are set to be   
$t^{\rm U}=0.003$ and $t^{\rm S}=0.024$. 
The spectra of MPM are shown in Fig.1 and denoted as curve 4.
The resulting ${\rm P_{MPM}}$ with value -0.68 is different from the 
${\rm P_{PM}}$ with value -0.37. 

\section{Numerical Result}

A recent experimental measurement of $\Lambda_b$ polarization, 
 $\rm P$, is determined through the variable $y$ proposed 
in \cite{Bonvicini-Randall} 
\bee
y=\fr{<E_l>}{<E_\nu>}\;\;, 
\eee
and the experimental measured quantity 
\bee
R=\fr{y(\rm P)}{y(0)} \;\;.
\eee 
$y$ could be expressed in terms of $\Lambda_b$ rest frame measurements 
in the form
\bee
y=\fr{<E_l^*>+<P_l^*({\rm P})>}{<E_\nu^*>+<P_\nu^*({\rm P})>}.
\label{y}
\eee
Since there are still no spectra available, the averaged
quantities are then theoretical dependent. 
We calculate the ${\rm P}$'s from the four 
formalisms and list their values in Table.1. We note that
${\rm P_{MQM}}$ with value -0.94 is very close to the $b$ quark 
polarization asymmetry, ${\rm A_{RL}}$, 
which was calculated at the $Z$ vertex in  
$Z\to b \bar{b}$ \cite{CLOSE} and expressed in the form 
\bee
{\rm A_{RL}}=-\fr{2 v_b a_b}{v_b^2 + a_b^2}=-0.936 \;\;, 
\label{PSM}
\eee
where $v_b$ and $a_b$ are the vector and axial vector couplings of the $b$
quark to the $Z$  boson, respectively.  
As shown in Table.2, the ratio of $<P_l^*>/<P_{\nu}^*>$ of MQM is 
about one half of that of QM.      
This leads to ${\rm P_{MQM}}$ larger by 75\% than ${\rm P_{QM}}$. 

The value of ${\rm P_{MPM}}$ , -0.68, in Table.1
is very close to the $\Lambda_b$ polarization, -0.68
estimated in \cite{FP}. The authors of \cite{FP} used
the HQEFT to  
estimate the polarization retention of the $\Lambda_b$ baryon
produced from the $b$ quark fragmentation processes. 
Another similar result of $\Lambda_b$ polarization was
calculated by employing the spectator diquark fragmentation model 
\cite{Saleev}.
Since the identification between the fragmentation function and 
the distribution function could be made in the infinite momentum
frame \cite{BAREISSPASCHOS}, which could be accessed under the
heavy quark limit.
There is, therefore, no surprise that our ${\rm P_{MPM}}$ matches 
the $\Lambda_b$ polarization determined from 
the fragmentation processes.   

\section{Conclusion}
 
In this paper we have constructed four formalisms based on
the factorization formula for $\Lambda_b \to X_q l \nb$.
We used the four formalisms to calculate their 
corresponding $\Lambda_b$ polarizations 
denoted as ${\rm P_{QM}}$, ${\rm P_{MQM}}$, ${\rm P_{PM}}$ 
and ${\rm P_{MPM}}$, respectively. The resulting ${\rm P_{MQM}}$
with value -0.94 is very close to the $b$ quark polarization
asymmetry with value -0.936. The ${\rm P_{MPM}}$ with value -0.68
is also close to the $\Lambda_b$ polarization with value -0.68.  
Our result is consistent with those estimates from the
Standard Model calculation of the $b$ quark polarization
asymmetry and the polarization retention of the $\Lambda_b$ baryon
produced from the $b$ quark fragmentation processes.

We have also developed an power expansion scheme which generalizes 
the naive collinear expansion scheme to include heavy massive quark partons
and to apply to decay processes.    

Finally, we emphasize the correctness of the measurement by the 
ALEPH Collaboration \cite{ALEPH}. Of course, there should need measuring
the spectra to make the measured $\Lambda_b$ polarization more consistent. 

\noindent
{\bf Acknowledgments:}
\hskip 0.6cm 
We would like to thank Hsiang-nan Li for many useful discussions 
and instructive suggestions. This work was supported by  
the National Science Council of R.O.C. under Grant No. NSC87-2811-M001-0031.

\appendix
\section{The collinear expansion}

We now derive the collinear expansion  skipped 
by the text. For the amplitude
\bee
M_a = \int \fr{d^4 k}{(2\pi)^4} S^{\mu\nu}(k)T(k)\;\;,
\eee
we apply the expansion $k=\xi p + (k-\xi p)$ to $S^{\mu\nu}(k)$
as in \yref{smunu} and recast $M_a$ in the form  
\bee
M_a \approx \int \fr{d^4 k}{(2\pi)^4} S^{\mu\nu}(k=\xi p)T(k) \;.      
\eee
It is easy to factorize $M_a$ as 
\bee
M_a&=&\int \fr{d^4 k}{(2\pi)^4}\left\{ [S^{\mu\nu}(k=\xi p)\s{P}_b]
[T(k) \fr{\s{n}}{4 P_b\cdot n}]\right. \nn  
&&+ [S^{\mu\nu}(k=\xi p)\s{P}_b][T(k) 
\fr{\s{p}}{4 P_b\cdot p}] \nn 
&&- [S^{\mu\nu}(k=\xi p)\s{S}_b\gamma_5][T(k) 
\fr{\s{n}\gamma_5}{4 S_b\cdot n}] \nn
&&\left.- [S^{\mu\nu}(k=\xi p)\s{S}_b\gamma_5][T(k) 
\fr{\s{p}\gamma_5}{4 S_b\cdot p}]\right\} \;\;,
\eee
where the bracket denotes the trace over Dirac indices.
We now prove that those terms involving the contraction of $T(k)$ 
with $\s{p}$ (or $\s{p}\gamma_5$)  
will lead to high order contributions. 
Recall that the $b$ quark propagator could be expanded in the form 
\bee
\fr{i}{\s{P}_b -M_b + i\epsilon}=i\fr{\hat{\s{P}}_b + M_b}
{\s{P}_b -M_b + i\epsilon}+ \fr{i\s{n}}{2 P_b\cdot n } \;. 
\label{sb}
\eee 
The contraction of $T(k)$ with $\s{p}$ 
is equivalent to the contraction of the $b$ quark propagator with $\s{p}$.
This leads to two effects:
(1) The contraction of the first term in \yref{sb} with $\s{p}$ 
leads to the form  
\bee
i\fr{\hat{\s{P}}_b + M_b}{\s{P}_b -M_b + i\epsilon}{\s{p}}
=\fr{i}{\s{P}_b -M_b + i\epsilon}[i(\s{k}-\xi\s{p})-i M_b]
\fr{i \s{n}}{2 P_b\cdot n }{\s{p}} \;\;.  
\label{contract1}
\eee
(2) For the contraction of the special propagator with $\s{p}$, 
we have the substitution  
\bee
\fr{i\s{n}}{2P_b\cdot n}\s{p} \longrightarrow 
\fr{i}{\s{P}_b -M_b +i\epsilon}(i\gamma_{\alpha})
\fr{i\s{n}}{2P_b\cdot n}\s{p}\;\;.  
\eee
By Fourier transforming $k-\xi p$ into $\omega^{\alpha}_{\alpha^{\prime}}
i\partial^{\alpha^{'}}$, the effect (1) leads to the expression
\bee
S^{\mu\nu}_{\alpha}(k,k)T^{\alpha}_1(k) 
\eee
with
\bee
T^{\alpha}_1(k)=\omega^{\alpha}_{\alpha^{\prime}}
\int d^4 x e^{i kx}\langle \Lambda_b|\bar{b}_v(0)
i\partial^{\alpha^{'}}b_v(x)|\Lambda_b\rangle\;\;, 
\eee
and the effect (2) induces the form 
\bee
S^{\mu\nu}_{\alpha}(k,k_1)T^{\alpha}_2(k,k1)
\eee
with
\bee
T^{\alpha}_2(k,k1)=\omega^{\alpha}_{\alpha^{'}}
\int d^4 x \int d^4 y 
e^{i(k-k_1)x}e^{ik_1 y} 
\langle \Lambda_b|\bar{b}_v (-g A^{\alpha^{'}}_a T^a)
(x)b_v(y)|\Lambda_b\rangle\;\;, 
\eee
where $\omega^{\alpha}_{\alpha^{'}}=g^{\alpha}_{\alpha^{'}}
-\bar{n}^{\alpha}n_{\alpha^{'}}$ and $n\cdot A=0$ gauge has been 
used. Summing the possible contributions, 
we express $S^{\mu\nu}_{\alpha}(k,k_1)$ and $S^{\mu\nu}_{\alpha}(k,k)$ as  
\bee
S^{\mu\nu}_{\alpha}(k,k)=S^{\mu\nu}_{\alpha}(k,k_1)&=& 
\mbox{Disc}[(i\gamma_{\alpha})\fr{i \s{n}}{2 P_b\cdot n }
\Gamma^{\mu}\fr{i\s{P}_q}{P_q^2+i\epsilon}\Gamma^{\nu}]\nn  
&&+\mbox{Disc}[
\Gamma^{\mu}\fr{i\s{P}_q}{P_q^2+i\epsilon}\Gamma^{\nu}
\fr{-i \s{n}}{2 P_b\cdot n }(-i\gamma_{\alpha})]\;\;.
\eee  
The mass $-i M_b$ involved term in \yref{contract1}
does not contribute because of the $V-A$ structure of the Standard Model.     
The cases for the contraction of $T(k)$   
with $\s{p}\gamma_5$ may be considered in the similar way.
As a result we arrive at 
\bee
M_a&=&\int \fr{d^4 k}{(2\pi)^4}\left\{ [S^{\mu\nu}(k)\s{P}_b]
[T(k) \fr{\s{n}}{4 P_b\cdot n}]\right. \nn  
&&\hspace{2cm}\left.- [S^{\mu\nu}(k)\s{S}_b\gamma_5][T(k)
\fr{\s{n}\gamma_5}{4 S_b\cdot n}] \right\}\nn
&&+ \int \fr{d^4 k}{(2\pi)^4} \int \fr{d^4 k_1}{(2\pi)^4}
\left\{[S^{\mu\nu}_{\alpha}(k,k_1)\s{P}_b][T^{\alpha}(k,k_1) 
\fr{\s{p}}{4 P_b\cdot p}]\right. \nn 
&&\hspace{2cm}\left.- [S^{\mu\nu}_{\alpha}(k,k_1)
\s{S}_b\gamma_5][T^{\alpha}(k,k_1) 
\fr{\s{p}\gamma_5}{4 S_b\cdot p}] \right\}\;\;,
\eee
where
\bee
T^{\alpha}(k,k1)&\equiv& T^{\alpha}_1(k,k)+ T^{\alpha}_2(k,k_1)\\\nn
&=&\omega^{\alpha}_{\alpha^{'}}\int d^4 x \int d^4 y
e^{i(k-k_1)x}e^{ik_1 y} \langle \Lambda_b|\bar{b}_v(0) iD^{\alpha^{'}}(x) 
(x)b_v(y)|\Lambda_b\rangle
\eee
with $iD^{\alpha}=i\partial^{\alpha}-g A^{\alpha}_a T^a$.

%

\newpage
Table.1 The values of the $\Lambda_b$ polarization 
 are determined from 
the quark model, the modified quark model, the parton model 
and the modified parton model. 
\vskip 1.0cm
\[\begin{array}{lccc} \hline\hline
{\rm P_{QM}} & {\rm P_{MQM}} & {\rm P_{PM}} & {\rm P_{MPM}}\\ \hline
-0.23 & -0.94 & -0.37 &-0.68\\
\hline\hline
\end{array} \]

Table.2 The values of the $<E_l^*>$, $<E_\nu^*>$, $<P_l^*>/{\rm P}$ 
and $<P_\nu^*>/{\rm P}$ are calculated from 
the quark model, the modified quark model, the parton model 
and the modified parton model. The units are in 1 $GeV$. 
\vskip 1.0cm
\[\begin{array}{lcccc} \hline\hline
{\rm Model}&<E_l^*>&<E_\nu^*>&<P_l^*>/{\rm P} &<P_\nu^*>/{\rm P} \\ \hline
{\rm QM}&0.3290 & 0.2820 & -0.0470&0.0940\\\hline 
{\rm MQM}&0.2731 &0.2379 &-0.0153 &0.0153 \\\hline 
{\rm PM}&0.1696 &0.1570 &-0.0242 &0.0238\\\hline
{\rm MPM}&0.1502 &0.1296 &-0.0105 &0.0124 \\
\hline\hline
\end{array} \]

\newpage
\begin{center}
{\large\bf Figure Captions}
\end{center}

\vskip 1.0cm
\noindent
{\bf Fig.1(a)}:\\
Unpolarized charged lepton spectra of the $\Lambda_b\to X_q\ell\nu$ 
decay are shown as curve (1) for the quark model, 
curve (2) for the modified quark model, curve (3) for 
the parton model, and curve (4) for the modified parton model.
\vskip 1.0cm
\noindent
{\bf Fig.1(b)}:\\
Polarized charged lepton spectra of the $\Lambda_b\to X_q\ell\nu$
decay are shown as curve (1) for the quark model,
curve (2) for the modified quark model, curve (3) for
the parton model, and curve (4) for the modified parton model. 
\vskip 1.0cm
\noindent
{\bf Fig.1(c)}:\\
Total charged lepton spectra of the $\Lambda_b\to X_q\ell\nu$
decay are shown as curve (1) for the quark model,
curve (2) for the modified quark model, curve (3) for
the parton model, and curve (4) for the modified parton model. 
The ${\rm P}\cos\theta_l=1$ condition has been used.
\newpage
\begin{figure}[htbp]
\epsfig{file=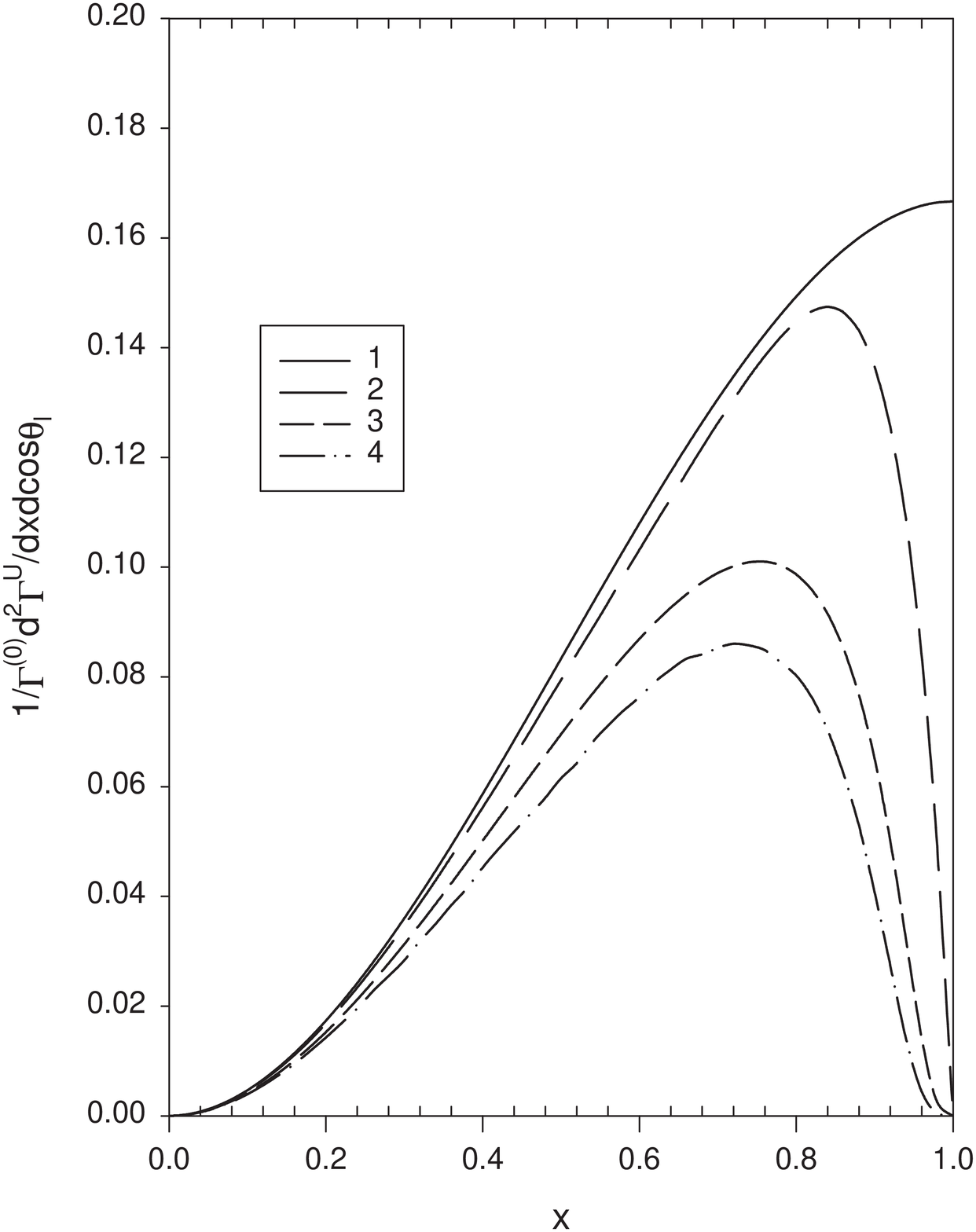,height=18cm,width=14cm}
\vskip 0.3cm
\begin{center}
Fig.1(a)
\end{center}
\end{figure}
\newpage
\begin{figure}[htbp]
\epsfig{file=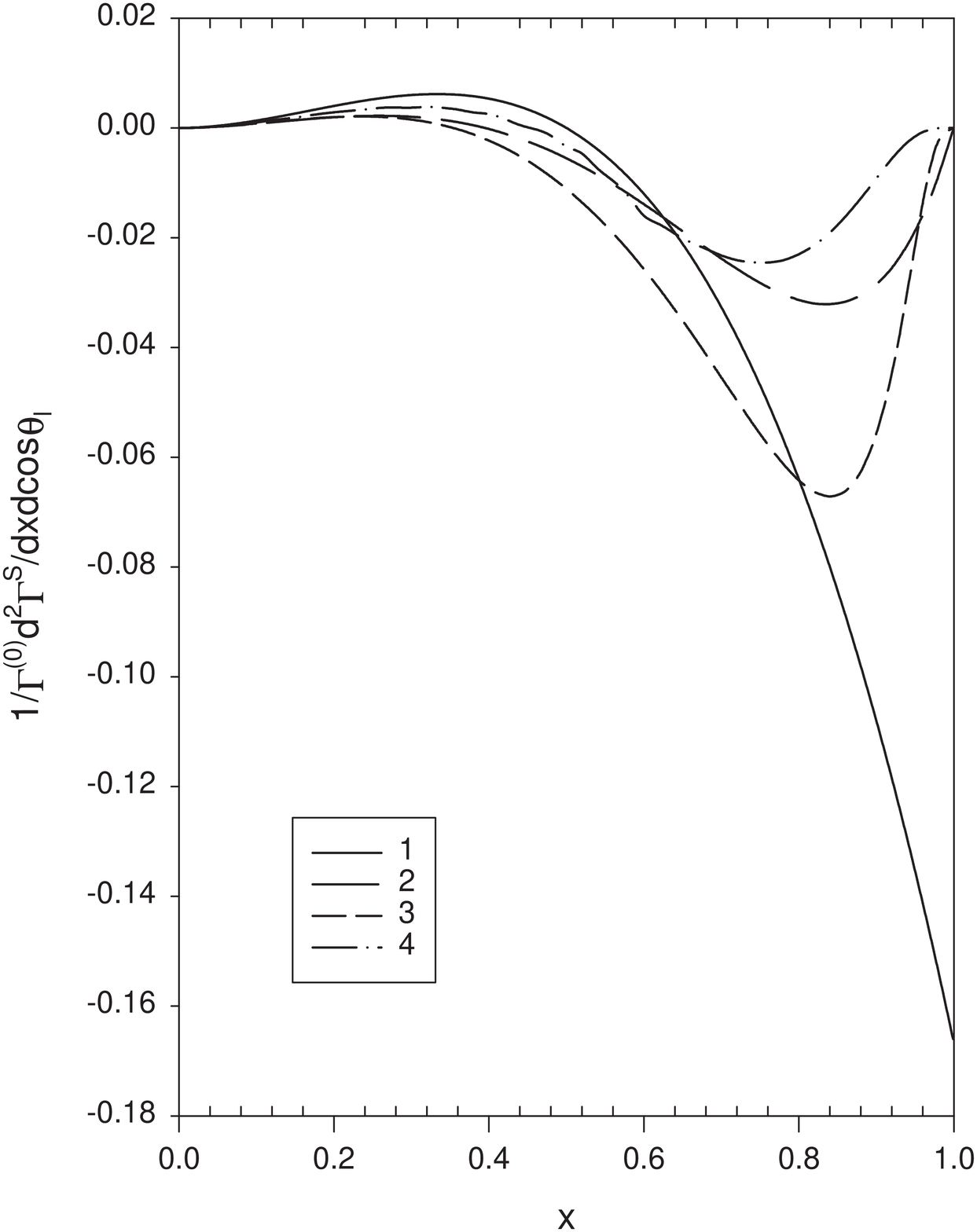,height=18cm,width=14cm}
\vskip 0.3cm
\begin{center}
Fig.1(b)
\end{center}
\end{figure}
\newpage
\begin{figure}[htbp]
\epsfig{file=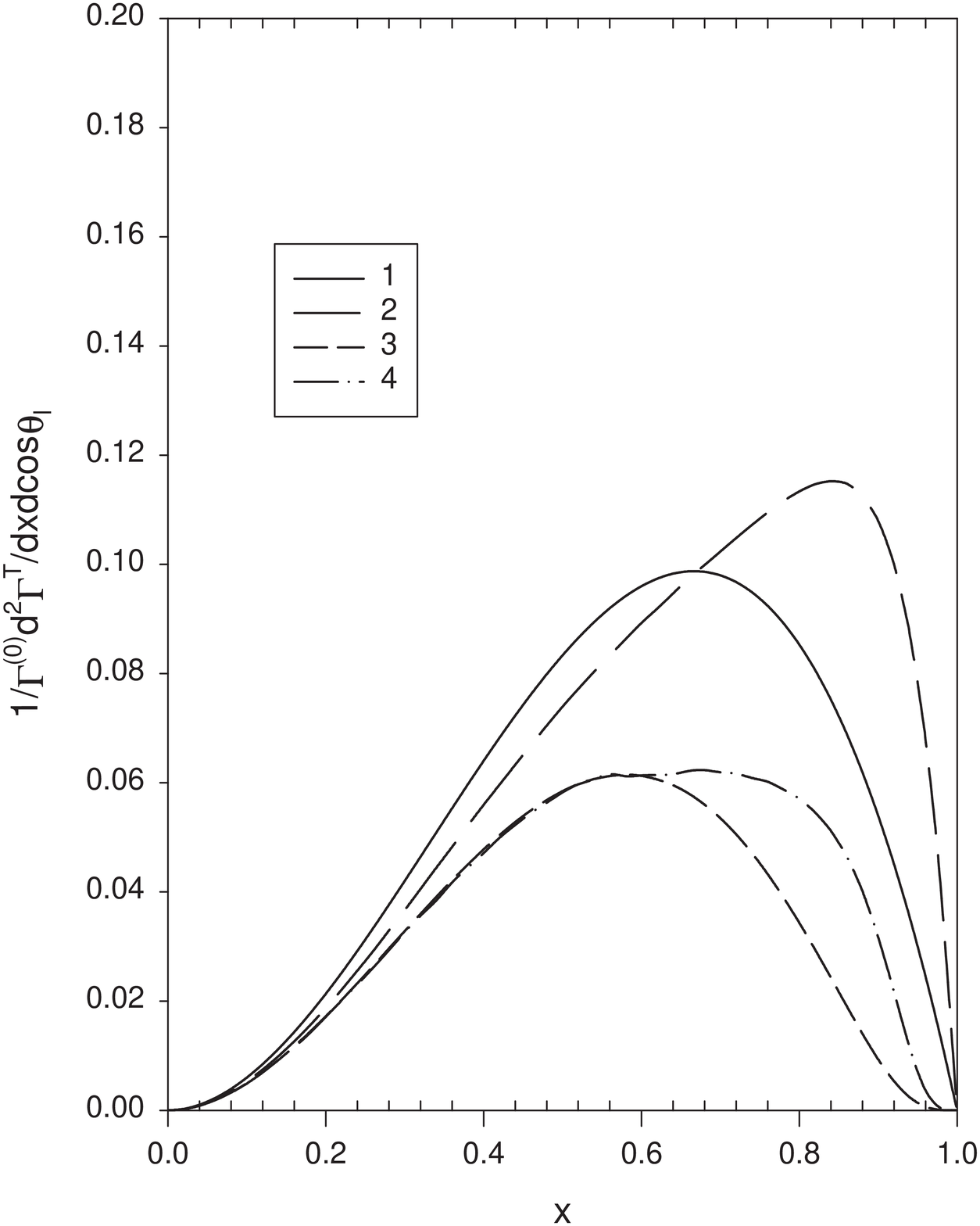,height=18cm,width=14cm}
\vskip 0.3cm
\begin{center}
Fig.1(c)
\end{center}
\end{figure}
\end{document}